\documentclass{PoS}
\usepackage[square, comma, sort&compress, numbers]{natbib}

\rightline{\parbox{4cm}{\large\rm
		ADP-17-3/T1009
	}}

\title{Study of Low-Lying Baryons with Hamiltonian Effective Field Theory}

\ShortTitle{Study of Low-Lying Baryons with Hamiltonian Effective Field Theory}

\author{\speaker{Zhan-Wei Liu}${}^{a,c}$, Jonathan M. M. Hall${}^a$, Waseem Kamleh${}^a$, Derek B. Leinweber${}^a$, Finn M. Stokes${}^a$, Anthony W. Thomas${}^{a,b}$, Jia-Jun Wu${}^a$\\
${}^a$Special Research Center for the Subatomic Structure of Matter (CSSM), Department of Physics,
University of Adelaide, Adelaide, South Australia 5005, Australia
\\
${}^b$ARC Centre of Excellence in Particle Physics at the Terascale, Department of Physics, University of Adelaide,
Adelaide, South Australia 5005, Australia\\
${}^c$School of Physical Science and Technology, Lanzhou University, Lanzhou 730000, China\\
E-mail: \email{liuzhanwei@lzu.edu.cn}}

\abstract{Drawing on experimental data for baryon resonances, Hamiltonian effective field theory (HEFT) is used to predict the positions of the finite-volume energy levels to be observed in lattice QCD simulations. We have studied the low-lying baryons $N^*(1535)$, $N^*(1440)$, and $\Lambda(1405)$. In the initial analysis, the phenomenological parameters of the Hamiltonian model are constrained by experiment and the finite-volume eigenstate energies are a prediction of the model. The agreement between HEFT predictions and lattice QCD results obtained at finite volume is excellent. These lattice results also admit a more conventional analysis where the low-energy coefficients are constrained by lattice QCD results, enabling a determination of resonance properties from lattice QCD itself. The role and importance of various components of the Hamiltonian model are examined in the finite volume. The analysis of the lattice QCD data can help us to undertand the structure of these states better.}

\FullConference{The 26th International Nuclear Physics Conference\\
		11-16 September, 2016\\
		Adelaide, Australia}

\begin{document}

\section{Introduction} \label{SecIntro}

The spectra and structures of hadrons are very important to the understanding of the strong interaction. To study them, many theories and models have been developed \cite{Kaiser1997,Oset1998,Oller2001,Ikeda2011,Thomas:1981vc,Thomas:1982kv,Doring:2011ip}. Much progress has been made, but there are still significant problems that remain unsolved.

Naive quark models predict that the mass of $N^*(1535)$ should be smaller than that of $N^*(1440)$ based on the assumption that these two nucleon excitations are made of three valence quarks. However, the mass of $N^*(1535)$ is larger. This contradiction indicates that the $\pi N$ and other two-particle states with a dominant five-quark component can play an important role in forming these excitations. Here, we carefully examine the effect of these two-particle states with Hamiltonian effective field theory (HEFT).

Lattice QCD is a first principles approach that yields non-perturbative calculations for the energy spectra and structures of hadronic states \cite{Menadue2012,Engel2013a,Kiratidis2015}. Like experimental scattering data, lattice QCD calculations can also provide key information on the properties of hadrons. HEFT can analyze both the lattice QCD data and experimental data at the same time to obtain valuable insight. It has been widely used in hadronic physics, with great success \cite{Wu2014,Matsuyama2007,Hall:2014uca,Liu2016a,Liu2016,Liu2016Lambda}.

In this talk, we use HEFT to study $N^*(1535)$, $N^*(1440)$, and $\Lambda(1405)$. The formalism is reviewed in Sec. \ref{SecFormalism}, and the results and discussions are listed in Secs. \ref{SecDisN1535}, \ref{SecDisN1440}, and \ref{SecDisLam1405}, respectively. We summarize in Sec. \ref{SecSummary}.

\section{Framework} \label{SecFormalism}
\subsection{Hamiltonian}
To study a baryon $|B\rangle$ with HEFT, one needs to know the interactions amongst the related particles. We use the following Hamiltonian to describe the interactions,
\begin{eqnarray}
H^B = H^B_0 + H^B_{\rm int}.
\label{eq:h}
\end{eqnarray} 
In the center-of-mass frame, the kinetic terms $H^B_0$ can be written 
\begin{eqnarray}
H^B_0 &=&\sum_{B_0} |B_0\rangle \, m_0^B  \, \langle B_0|+ \sum_{\alpha}\int d^3\vec{k}\,
|\alpha(\vec{k})\rangle\, \left[\, \omega_{\alpha_M}(k)+\omega_{\alpha_B}(k)\, \right]
\,\langle\alpha(\vec{k})| \, ,
\label{eq:h0}
\end{eqnarray}
where $|B_0\rangle$ is a bare baryon and $|\alpha(\vec k)\rangle$ are the two-particle states with the same quantum numbers as the baryon $|B\rangle$. In the case of the $N^*(1535)$, the two-particle states $|\alpha\rangle$ can be $|\pi N\rangle$, $|\pi \eta\rangle$, and so on \cite{Liu2016a}. For the $N^*(1440)$ and $\Lambda(1405)$, refer to Refs.~\cite{Liu2016,Liu2016Lambda} for details. $m_0^B$ is the bare mass, while $\omega_{\alpha_M}(k)$ and $\omega_{\alpha_B}(k)$ are the kinetic energies of the meson and baryon in the state $|\alpha(\vec k)\rangle$,
$\omega_X(k)=\sqrt{m_X^2+k^2}.$

The interaction Hamiltonian $H^B_{\rm int}$ can be divided into two parts
\begin{eqnarray}
H^B_{\rm int} = g^B + v^B.\label{eq:hi}
\end{eqnarray}
$g^B$ describes the interaction between the bare baryon and the two-particle states
\begin{eqnarray}
g^B &=& \sum_{\alpha, B_0}\int d^3\vec{k} \, \left\{\,  |\alpha(\vec{k})\rangle \,
G^{B\dagger}_{\alpha, B_0}(k) \, \langle B_0| 
+
|B_0\rangle \, G^B_{\alpha, B_0}(k) \, \langle \alpha(\vec{k})|\, \right\},
\label{eq:int-g}
\end{eqnarray}
and $v^B$ describes the direct two-to-two particle interactions
\begin{eqnarray}
v^B = \sum_{\alpha,\beta} \int d^3\vec{k} \, d^3\vec{k}'\, |\alpha(\vec{k})\rangle \,
V^{B}_{\alpha,\beta}(k,k') \, \langle \beta(\vec{k}')|\, .
\label{eq:int-v}
\end{eqnarray}
The detailed forms of $G^B_{\alpha, B_0}(k)$ and $V^{B}_{\alpha,\beta}(k,k')$ can be found in Refs. \cite{Liu2016a,Liu2016,Liu2016Lambda}.

\subsection{$T$-matrix at infinite volume}
We can obtain the $T$-matrix by solving a three-dimensional reduction of the Bethe-Salpeter equation in the infinite volume,
\begin{eqnarray}
&&T^B_{\alpha, \beta}(k,k';E)=\tilde V^B_{\alpha, \beta}(k,k';E)+\sum_\gamma \int q^2 \, dq\, \tilde V^B_{\alpha, \gamma}(k,q;E) \, \frac{1}{E-\omega_\gamma(q)+i \epsilon} \,  T^B_{\gamma, \beta}(q,k';E),
\label{eq:BS}
\end{eqnarray}
where $\omega_\gamma(q)$ is the energy of the two-particle state $|\gamma(\vec k)\rangle$, and the coupled-channel potential can be obtained from the interaction Hamiltonian 
\begin{eqnarray}
\tilde V^ B_{\alpha, \beta}(k,k';E) &=& G^{B\dag}_{\alpha, B_0}(k) \, \frac{1}{E-m_0^B+i\epsilon} \, G^B_{\beta, B_0}(k') 
+V^B_{\alpha,\beta}(k,k').
\label{eq:lseq-2}
\end{eqnarray}
Using the $T$-matrix one can easily extract the phaseshifts, inelasticities, cross sections, and so on.

\subsection{Finite-volume matrix Hamiltonian model}
In the finite volume particles can only carry a discretized momenta, $k_n=2\pi \sqrt{n}/L,$ where $n = n_x^2 + n_y^2 + n_z^2$ is an integer representing the momentum magnitude and $L$ is the length of the box. We first need to discretize the Hamiltonian in the finite volume. Taking the $N^*(1535)$ as an example, the non-interacting Hamiltonian is
\begin{equation}
\mathcal H_0^B={\rm diag}\left \{ m_0^B,\, \omega_{\pi N}(k_0),\, \omega_{\eta
	N}(k_0),\, \omega_{\pi N}(k_1),\, \omega_{\eta N}(k_1), \ldots
\right \} \, .
\end{equation}
The associated interaction Hamiltonian is
\begin{equation}
\mathcal H_I^B=\left( \begin{array}{cccccc}
0&\mathcal G^B_{\pi N,B_0}(k_0)&\mathcal G^B_{\eta N,B_0}(k_0)&\mathcal G^B_{\pi N,B_0}(k_1)&\mathcal G^B_{\eta N,B_0}(k_1)&\ldots\\
\mathcal G^B_{\pi N,B_0}(k_0)&\mathcal V^B_{\pi N,\pi N}(k_0,k_0)&\mathcal V^B_{\pi N,\eta N}(k_0,k_0)&\mathcal V^B_{\pi N,\pi N}(k_0,k_1)&\mathcal V^B_{\pi N,\eta N}(k_0,k_1)&\ldots\\
\mathcal G^B_{\eta N,B_0}(k_0)&\mathcal V^B_{\eta N,\pi N}(k_0,k_0)&\mathcal V^B_{\eta N,\eta N}(k_0,k_0)&\mathcal V^B_{\eta N,\pi N}(k_0,k_1)&\mathcal V^B_{\eta N,\eta N}(k_0,k_1)&\ldots\\
\mathcal G^B_{\pi N,B_0}(k_1)&\mathcal V^B_{\pi N,\pi N}(k_1,k_0)&\mathcal V^B_{\pi N,\eta N}(k_1,k_0)&\mathcal V^B_{\pi N,\pi N}(k_1,k_1)&\mathcal V^B_{\pi N,\eta N}(k_1,k_1)&\ldots\\
\mathcal G^B_{\eta N,B_0}(k_1)&\mathcal V^B_{\eta N,\pi N}(k_1,k_0)&\mathcal V^B_{\eta N,\eta N}(k_1,k_0)&\mathcal V^B_{\eta N,\pi N}(k_1,k_1)&\mathcal V^B_{\eta N,\eta N}(k_1,k_1)&\ldots\\
\vdots&\vdots&\vdots&\vdots&\vdots&\ddots\\
\end{array}
\right),
\label{eq:intHamMat}
\end{equation}
where
\begin{eqnarray*}
\mathcal G^B_{\alpha,B_0}(k_n)=\sqrt{\frac{C_3(n)}{4\pi}}  \left(\frac{2\pi}{L}\right)^{3/2} \, G^B_{\alpha,B_0}(k_n),~{\rm and}~ 
\mathcal V^B_{\alpha,\beta}(k_n, k_m)=\frac{\sqrt{C_3(n)\,C_3(m)}}{4\pi} \left(\frac{2\pi}{L}\right)^{3} \,
V^B_{\alpha,\beta}(k_n, k_m) .
\end{eqnarray*}
$C_3(n)$ represents the degeneracy factor for summing the squares of three integers to equal $n$.

As the pion mass varies, the masses of other hadrons will also change. For the mass of the bare baryon, we use 
\begin{equation}
	m_0^B(m_\pi^2)=m_0^B\mid_{\rm phys.}+\,\,\alpha_0^B\,(m_\pi^2-m_\pi^2\mid_{\rm phys.}).
\end{equation}
The eigenvalues of the discretized Hamiltonian provide the spectrum in the finite volume, and they can be used to analyze the lattice QCD data.

\section{Numerical results and discussion for $N^*(1535)$} \label{SecDisN1535}

HEFT can be used to connect the experimental data and the lattice QCD results. We fit the parameters in the Hamiltonian to the phaseshifts and inelasticities of $\pi N$ scattering in Sec.~\ref{SecPhaseshift}. In Sec.~\ref{SecFinite}, we give the predictions for the finite-volume spectrum from the fit parameters and make a comparison with lattice QCD results. In Sec. \ref{SecExtract}, we extract the pole for $N^*(1535)$ at infinite volume from the data of lattice QCD with HEFT. 

\subsection{Phaseshifts and inelasticities} \label{SecPhaseshift}

\begin{figure}[h]
\begin{center}
\scalebox{0.4}{\includegraphics{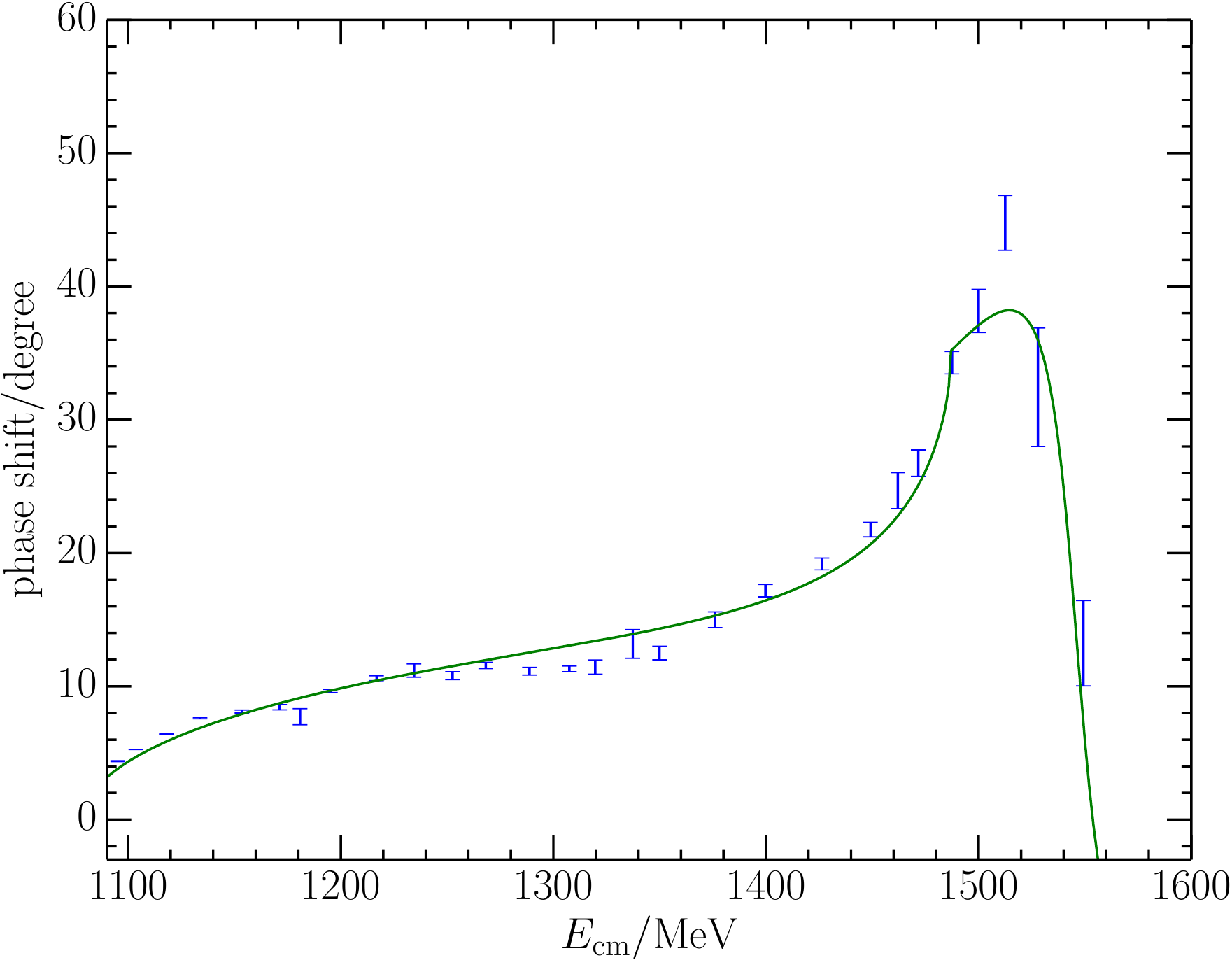}}\quad
\scalebox{0.4}{\includegraphics{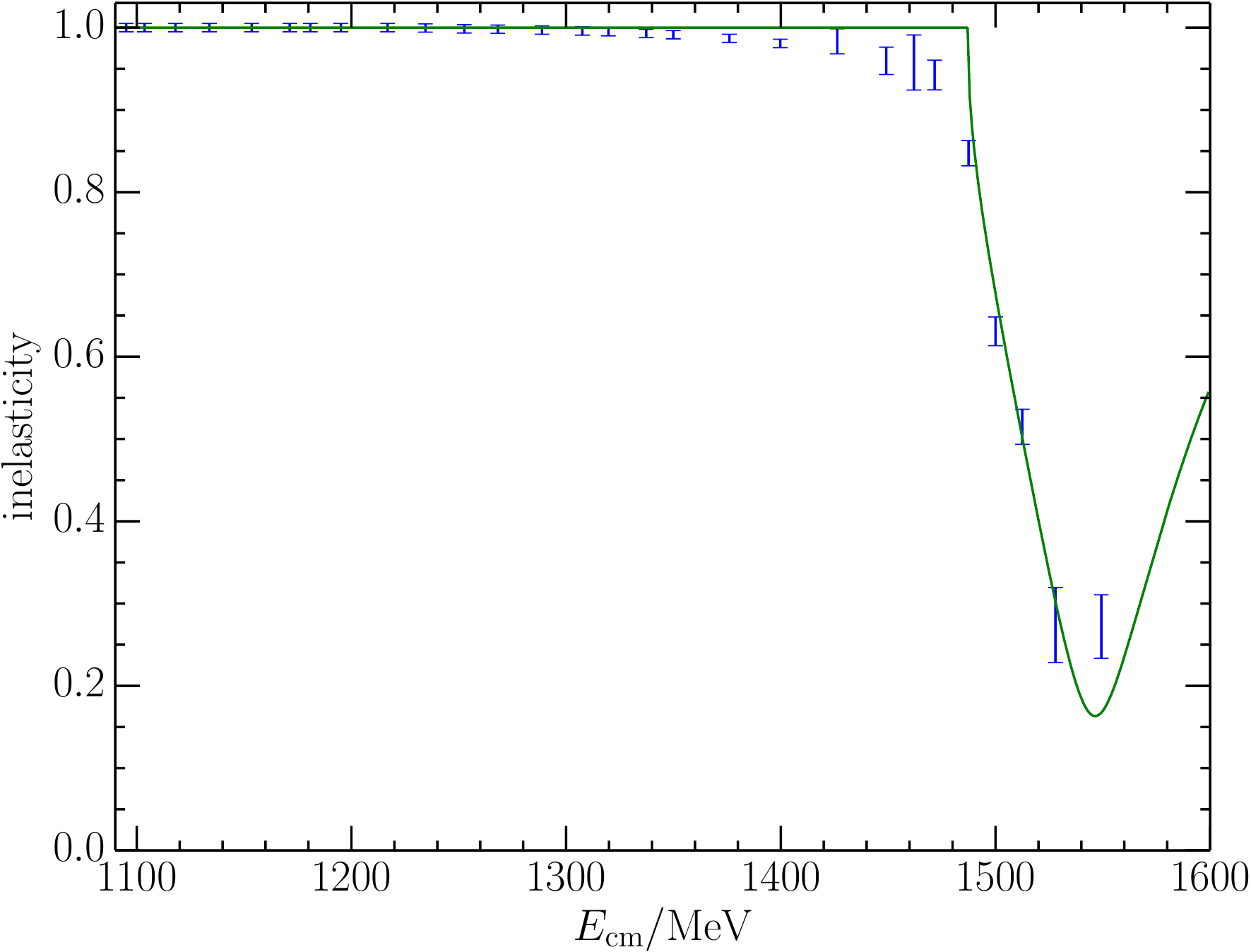}}
\caption{{\bf Color online:} Phaseshifts (left) and inelasticities (right) for $\pi N$ scattering with $I(J^P)=1/2(1/2^-)$ and $S=0$.}
\label{fig:N1535PSEta}
\end{center}
\end{figure}

Here we consider the interactions between the bare $N^*(1535)$, $\pi N$, and $\eta N$ states. The $\pi N$-$\pi N$ interaction is very important to the phaseshifts at low energies. We show our fits to the phaseshifts and inelasticities in Fig.~\ref{fig:N1535PSEta}. The model describes the experimental data well. Based on the fit parameters, we find a pole for the $N^*(1535)$ at $1531\pm 29-i 88\pm 2$ MeV.

\subsection{Finite-volume results} \label{SecFinite}
\begin{figure}[p]
\vspace{-0.4em}
\begin{center}
\scalebox{0.4}{\includegraphics{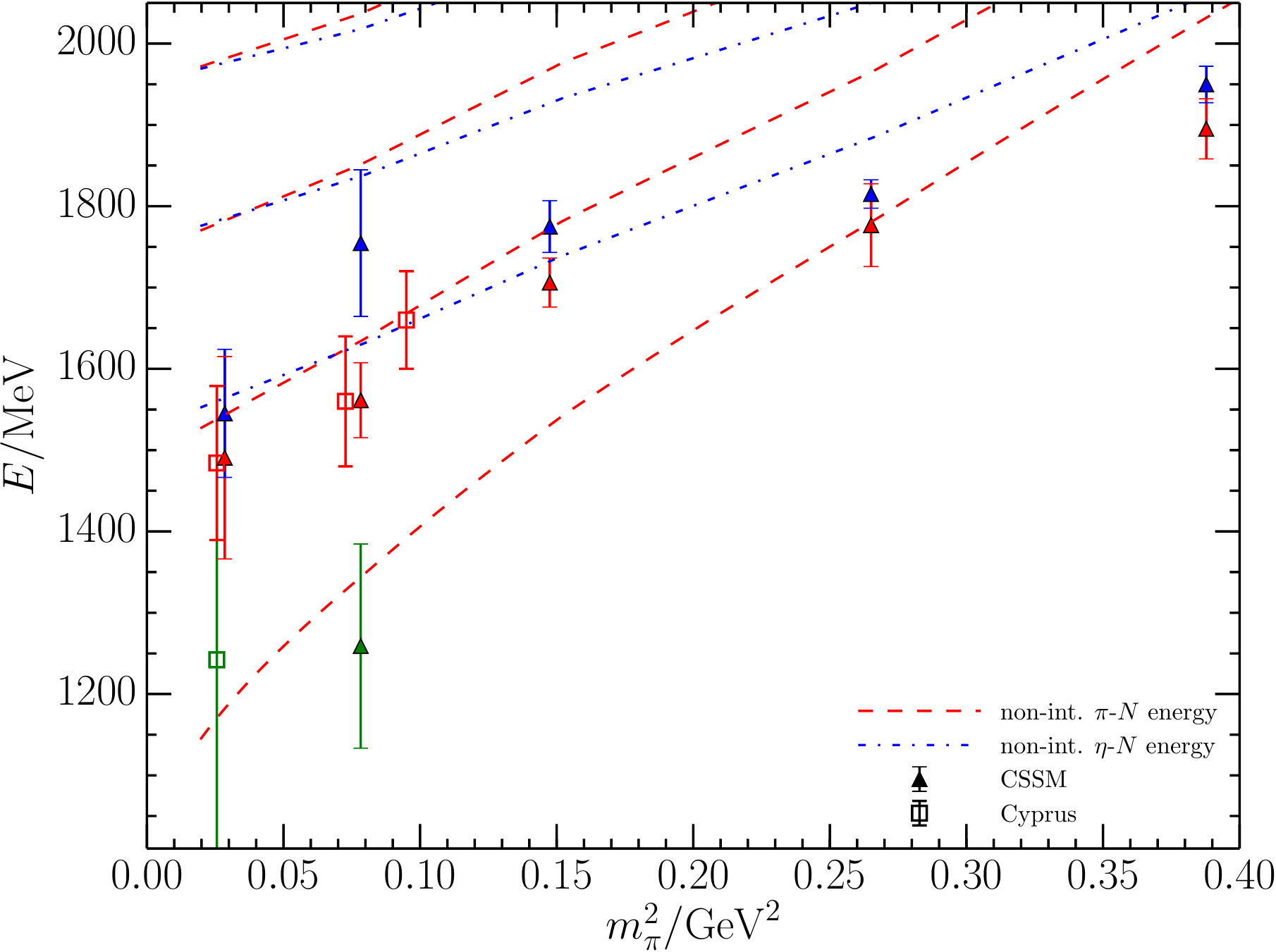}}
\scalebox{0.4}{\includegraphics{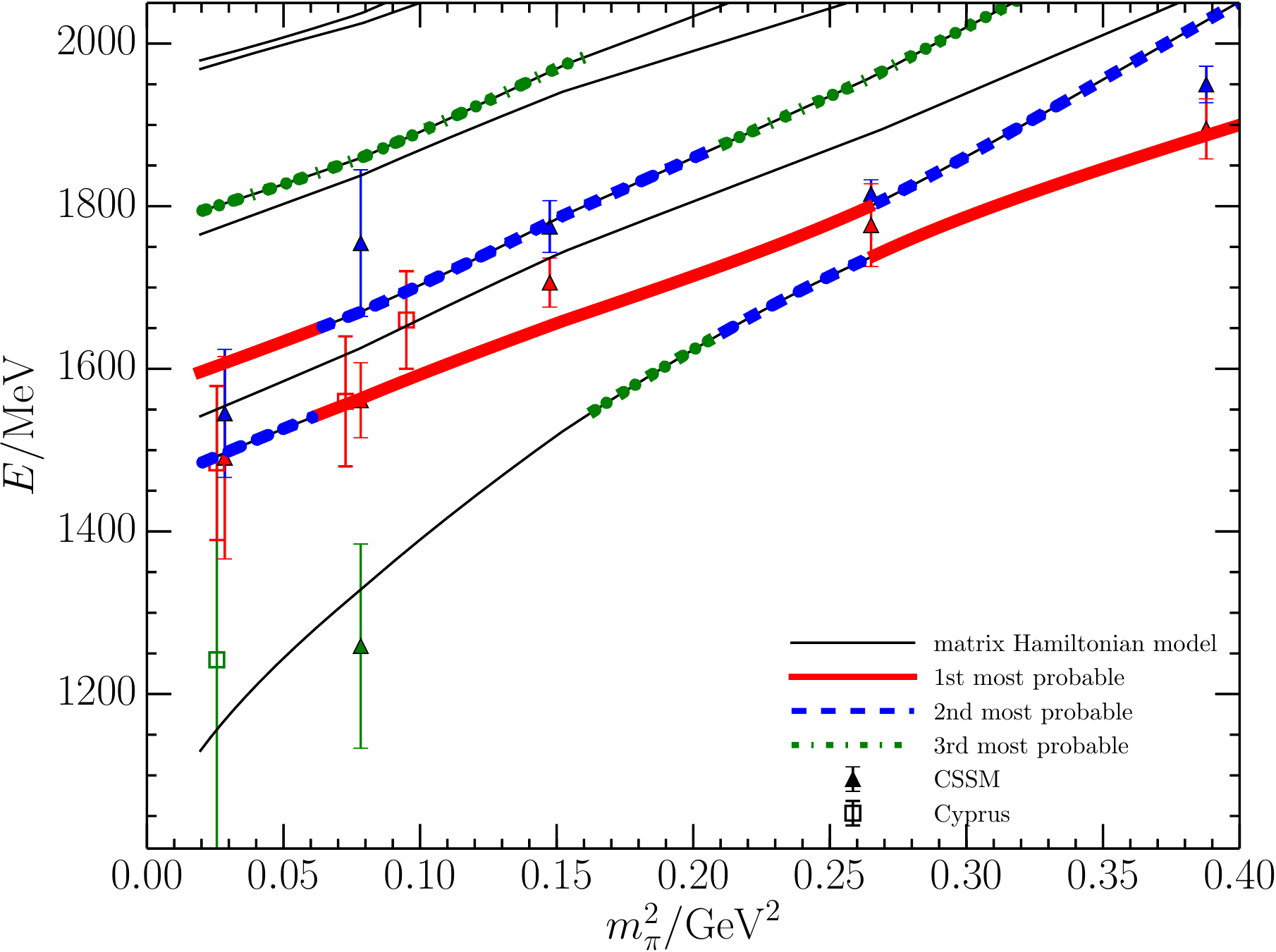}}
\caption{{\bf Color online:} The non-interacting energies of the low-lying two-particle meson-baryon channels (left) and the energy spectrum of HEFT (right) in the finite volume with $L\approx3$ fm for $I(J^P)=1/2(1/2^-)$ and $S=0$. The data
 with filled symbols are from the CSSM group, and those with hollow symbols are from the Cyprus group. The different line types and colors used in illustrating the energy levels indicate the strength of the bare basis state in the Hamiltonian-model eigenvector describing the composition of the state. The thick-solid (red), dashed (blue) and dotted (green) lines correspond to the states having the first, second, and third largest bare-state contributions, and therefore the most likely states to be observed with three-quark interpolating fields.}
%\label{fig:N1535Spectrum}
\end{center} 
\end{figure}

\begin{figure}[p]
\vspace{-0.3em}
\begin{center}
\scalebox{0.37}{\includegraphics{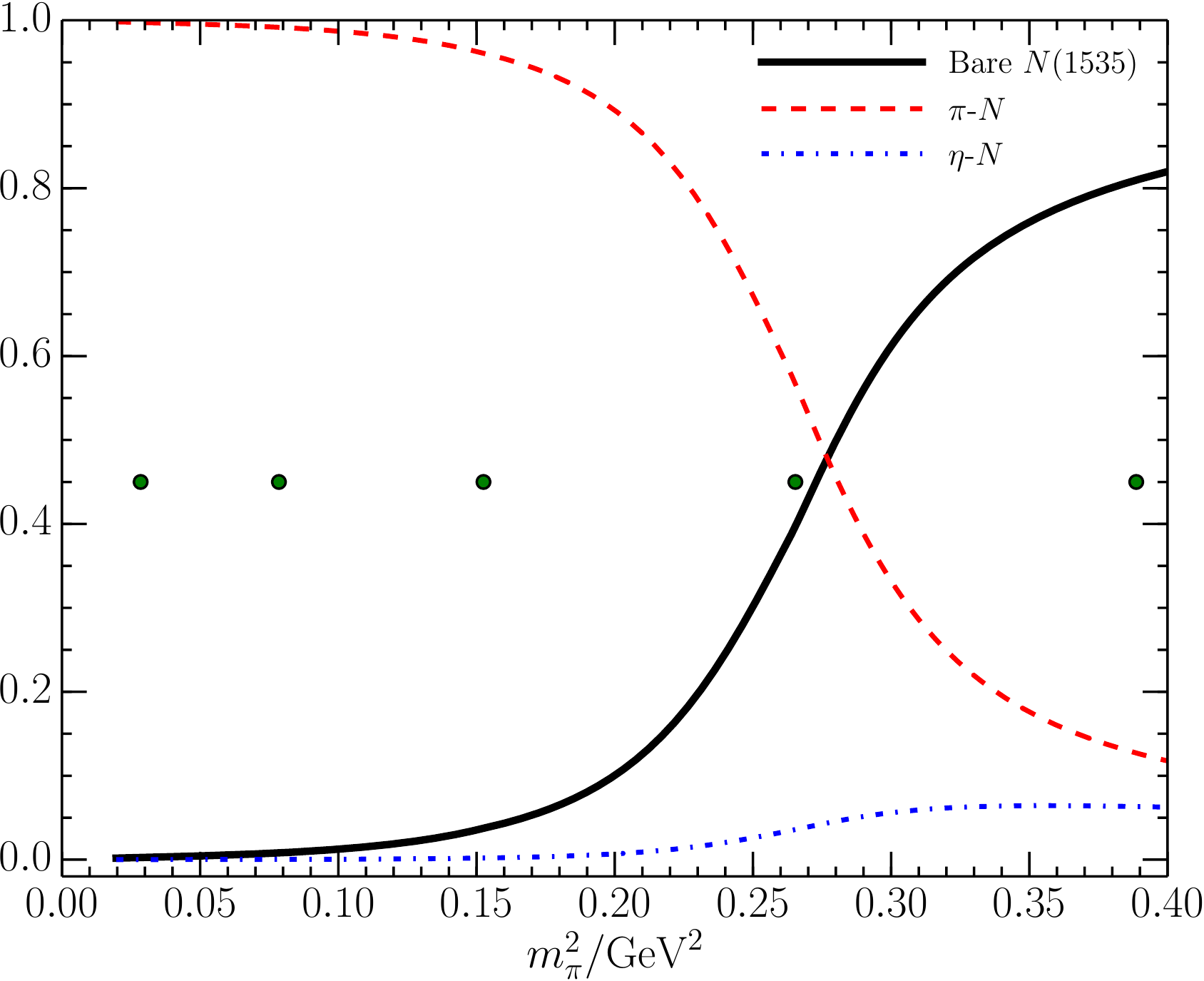}}
\scalebox{0.37}{\includegraphics{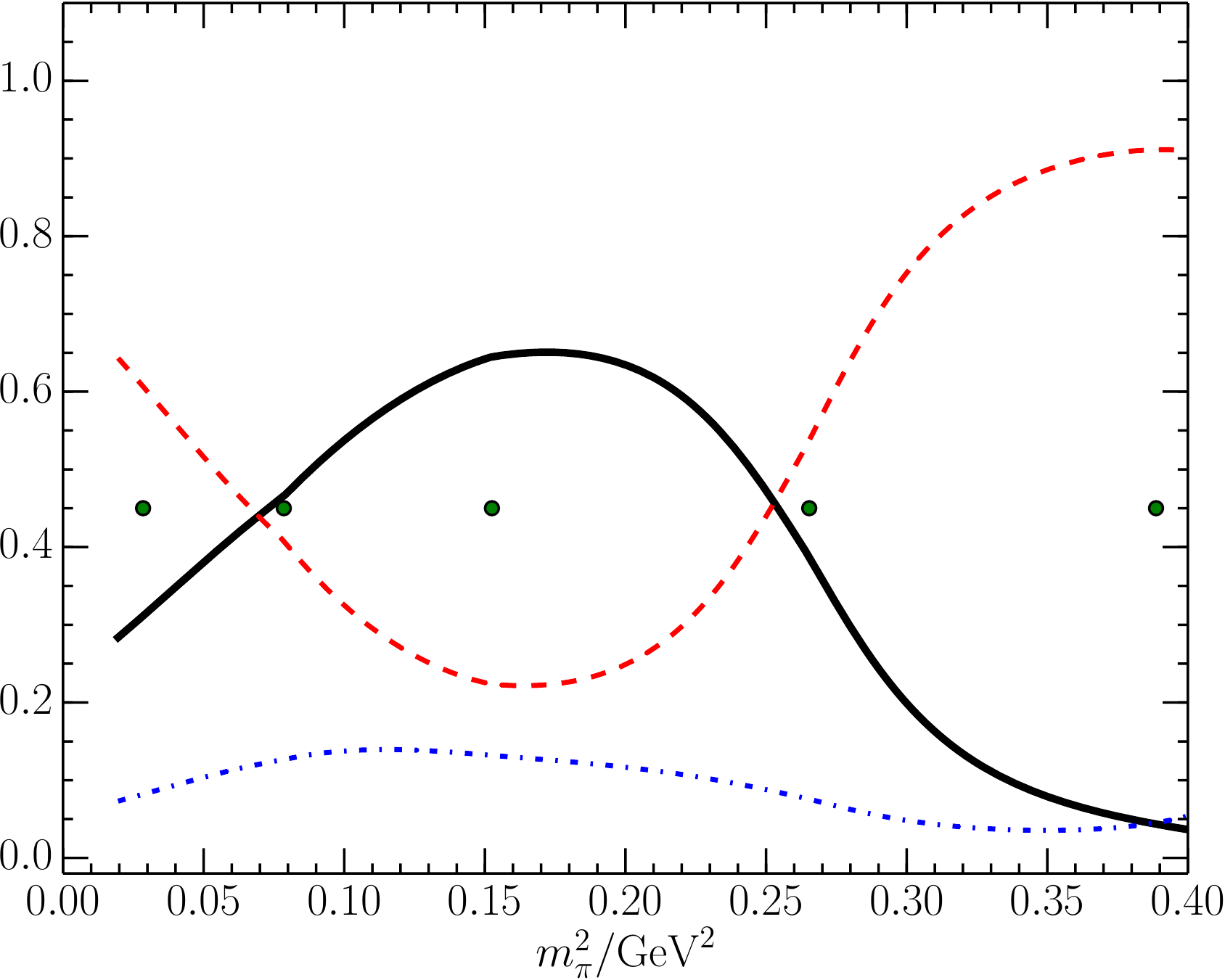}}
\scalebox{0.37}{\includegraphics{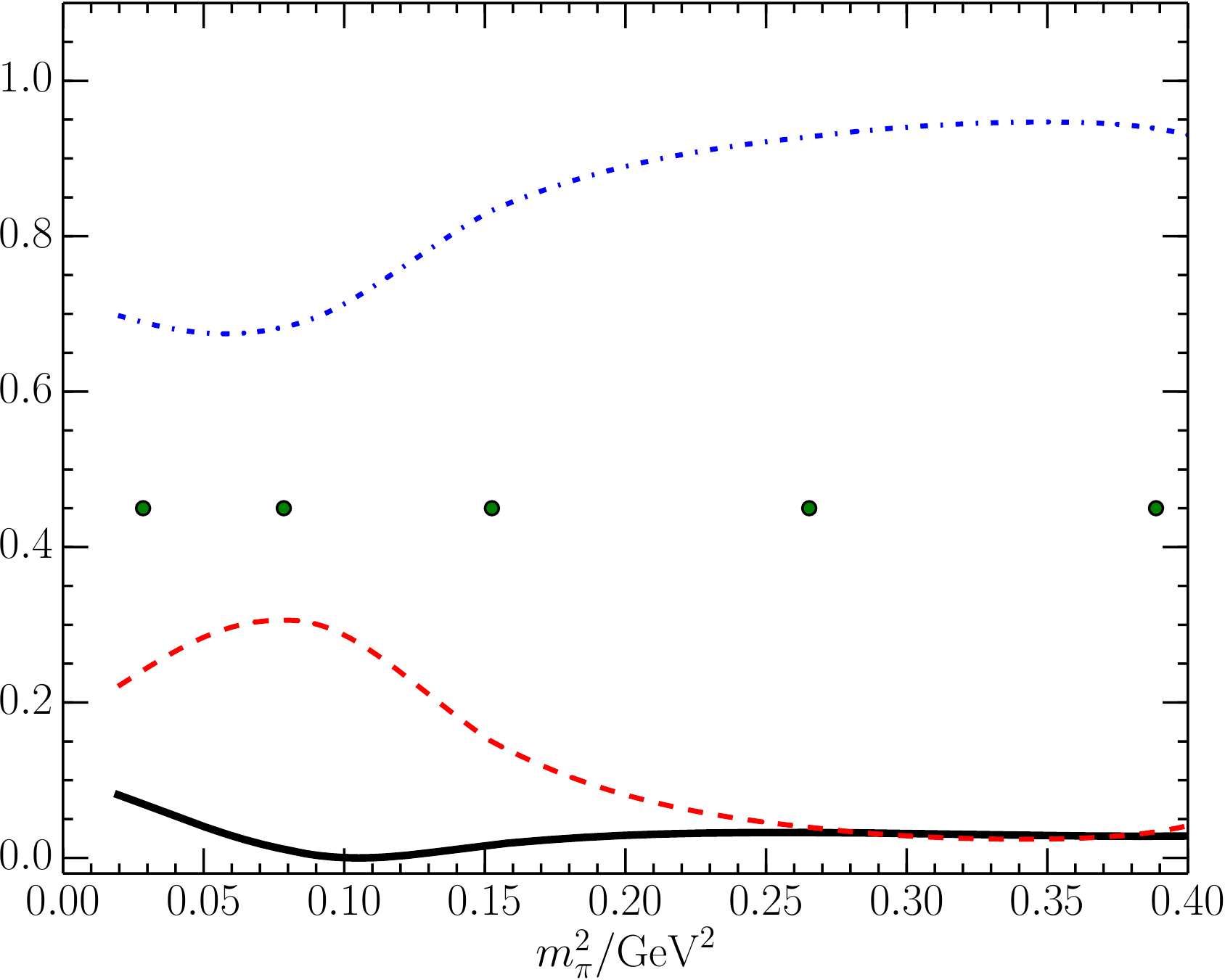}}
\scalebox{0.37}{\includegraphics{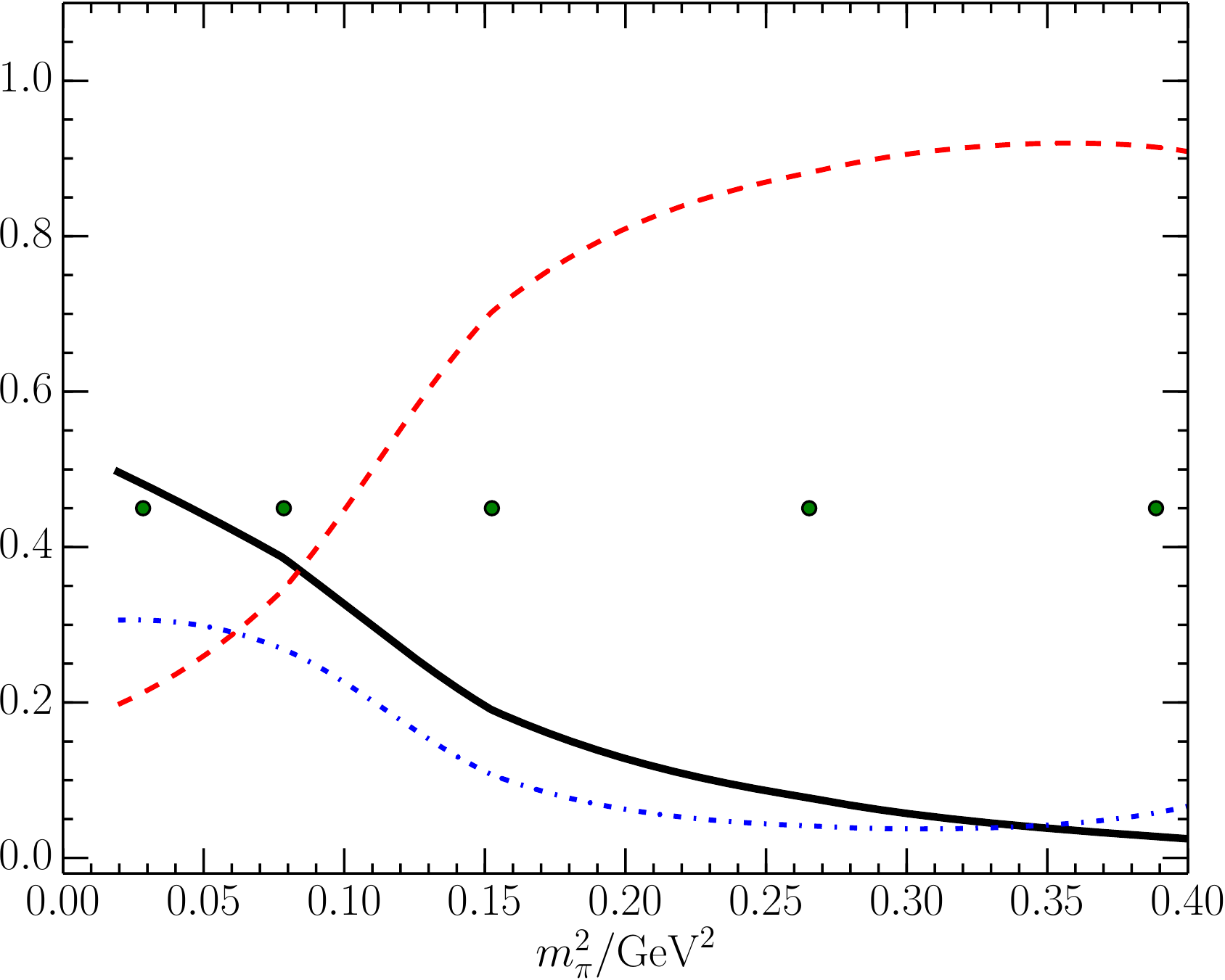}}
\caption{{\bf Color online:} The pion-mass evolution of the
Hamiltonian eigenvector components for $I(J^P)=\frac12(\frac12^-)$ and $S=0$. The top-left, top-right, bottom-left, and bottom-right graphs represent the 1st, 2nd, 3rd, and 4th eigenstates, respectively.}
\label{fig:N1535Structure}
\end{center} 
\end{figure}

With the parameters fit by the experimental data, we can study the effect of the interactions on the finite-volume spectrum. We list the energy levels without (left) and with (right) interactions in a box with length of about 3 fm in Fig.~\ref{fig:N1535Spectrum}. The lattice QCD data are also shown. We note that the interactions among the related states are critical to the consistency between our model prediction and the lattice QCD data.

Usually lattice QCD groups use local three-quark interpolators to extract the signals, and thus the eigenstates with a significant bare baryon component should be easier to observe on the lattice, since the coupling to dominant $\pi N$ or $\pi \eta$ multi-particle states are volume suppressed. We have colored the most probable eigenstates to be observed in the right graph of Fig. \ref{fig:N1535Spectrum}. One can see the lattice QCD data does preference the colored lines.

We can also analyze the structure of the eigenstates in the finite volume. We list the components for the first four eigenstates with HEFT in Fig. \ref{fig:N1535Structure}. From the top-left subfigure in Fig. \ref{fig:N1535Structure}, we notice the first eigenstate is mainly $\pi N$ scattering states at small pion masses, while it tends to be dominated by the bare state at large pion masses. The second eigenstate is a mix of the bare baryon and $\pi N$ scattering states, while the third eigenstate is dominated by the $\eta N$ states. The fourth eigenstate at small pion mass is a nontrivial mix of the bare state, $\pi N$, and $\eta N$ states.

\subsection{Information extracted by the lattice data} \label{SecExtract}

In the previous subsections, we obtain the bare mass $m_0^B$ by fitting the scattering data at infinite volume, and then use $m_0^B$ to see what happens in the finite volume. We do the reverse in this subsection, adjusting $m_0^B$ in order to fit the lattice QCD data. With this method, we obtain a pole at $1563^{+52}_{-80}-i 89.2^{+0.2}_{-4.2}$ MeV for the $N^*(1535)$ state in the infinite volume.

\section{Numerical results and discussion for $N^*(1440)$} \label{SecDisN1440}
The structure of $N^*(1440)$ is still under debate. Some models include a three-quark core, but the experimental data can also be explained under the assumption that this resonance is dynamically generated by the interplay of $\pi N$, $\pi \Delta$, and other two-particle states. 

\begin{figure}[htbp]
\begin{center}
\includegraphics[width=0.49\columnwidth]{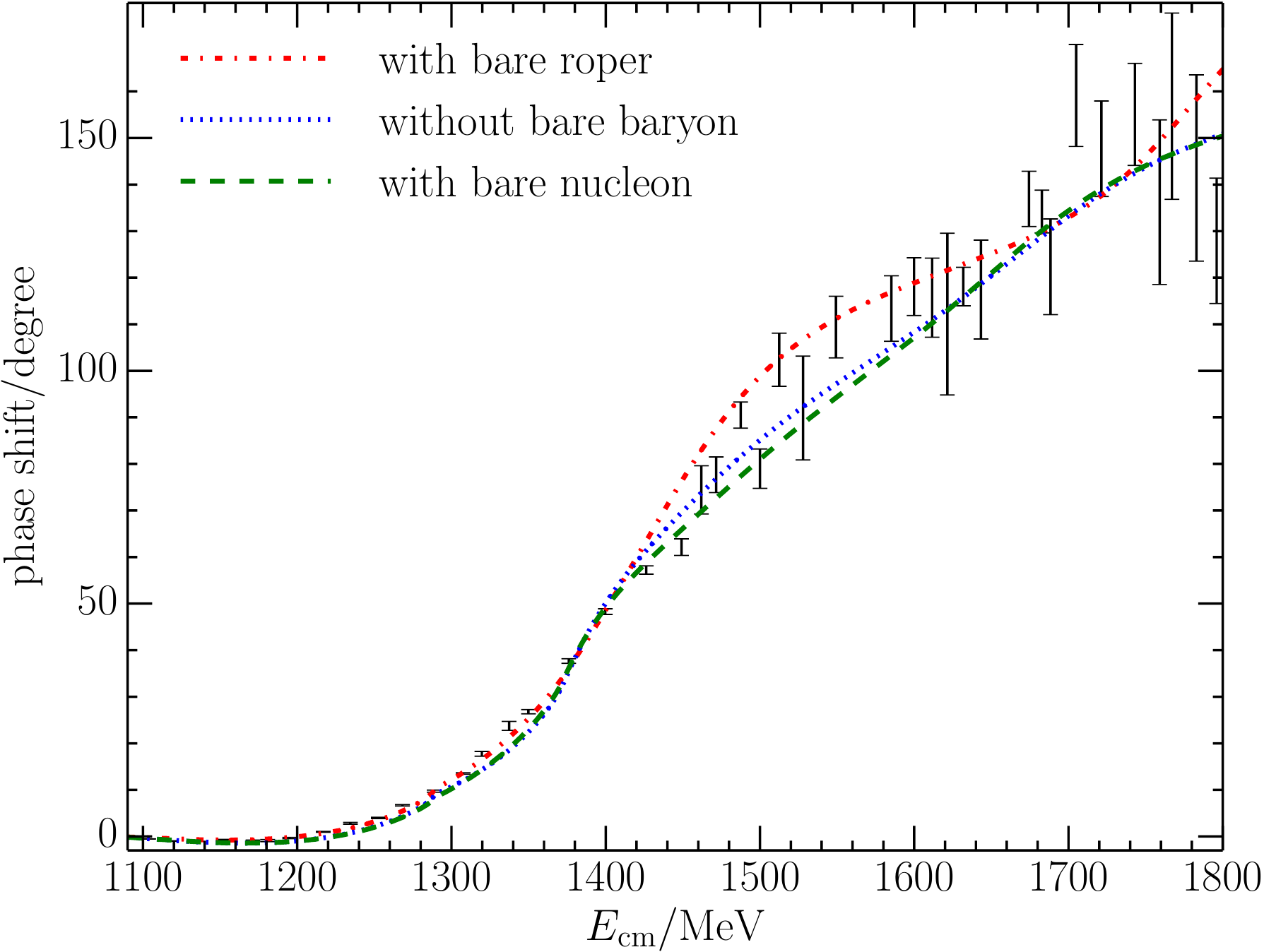}
\includegraphics[width=0.49\columnwidth]{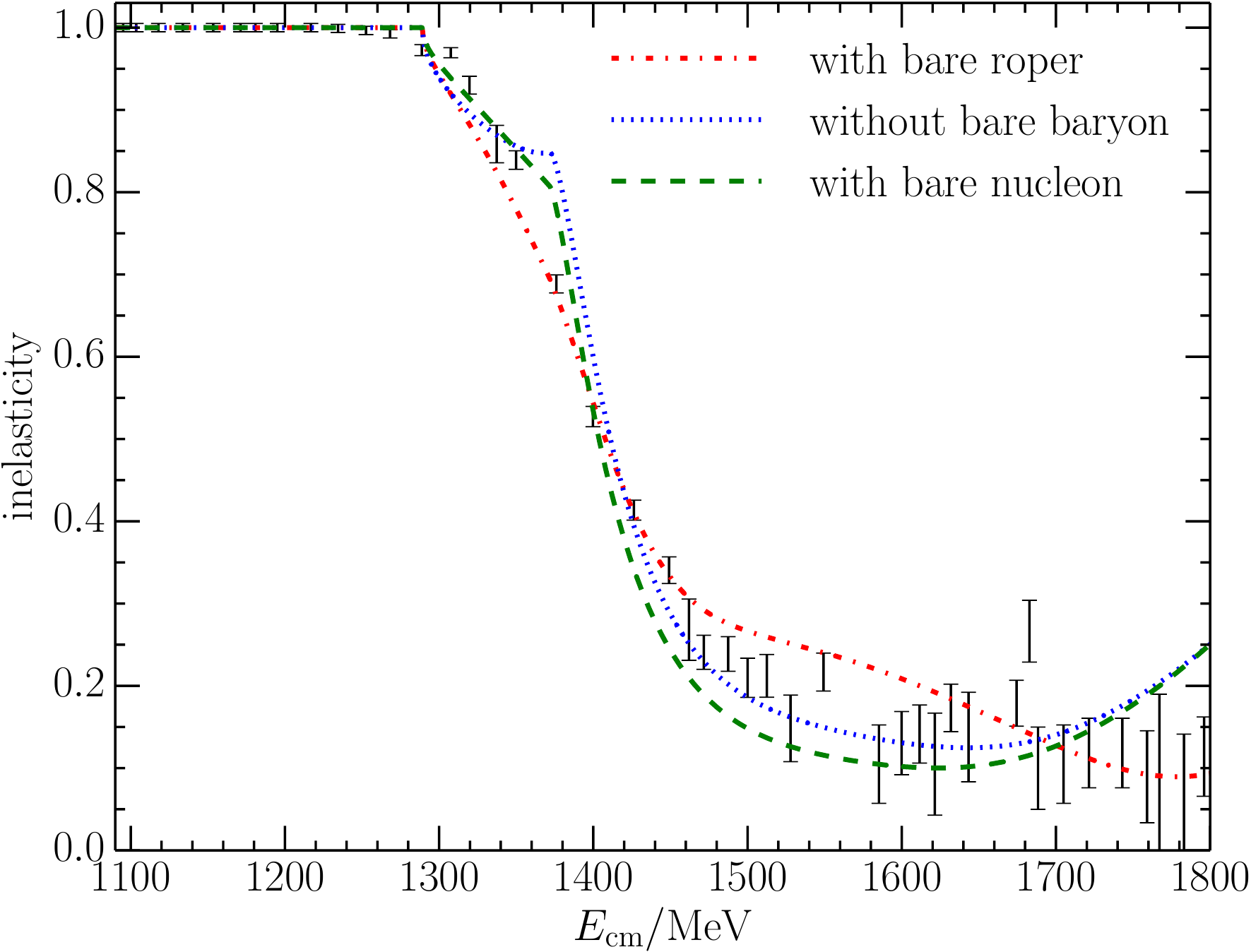}
\caption{{\bf Color online:} Phaseshifts (left) and inelasticities (right) for $\pi N$ scattering with $I(J^P)=\frac12(\frac12^+)$ and $S=0$.  The dot-dashed, dotted and dashed lines represent our best fits for scenario I with the bare $N^*(1440)$ (Roper), scenario II without a bare baryon, and scenario III with the bare nucleon, respectively.}\label{figPSEta}
\end{center} 
\end{figure}

We have considered three scenarios for the structure of the $N^*(1440)$. The first assumes that the $N^*(1440)$ contains a three-quark core, while the second scenario postulates that this resonance is purely dynamically generated by two-particle states. The third scenario is based on the second one, but also including corrections from a bare nucleon component. The fit for the phaseshifts and inelasticities for these three scenarios is shown in Fig.~\ref{figPSEta}, and we can see all these three scenarios can explain the experimental data.

\begin{figure}[tbp]
\begin{center}
\includegraphics[width=0.49\columnwidth]{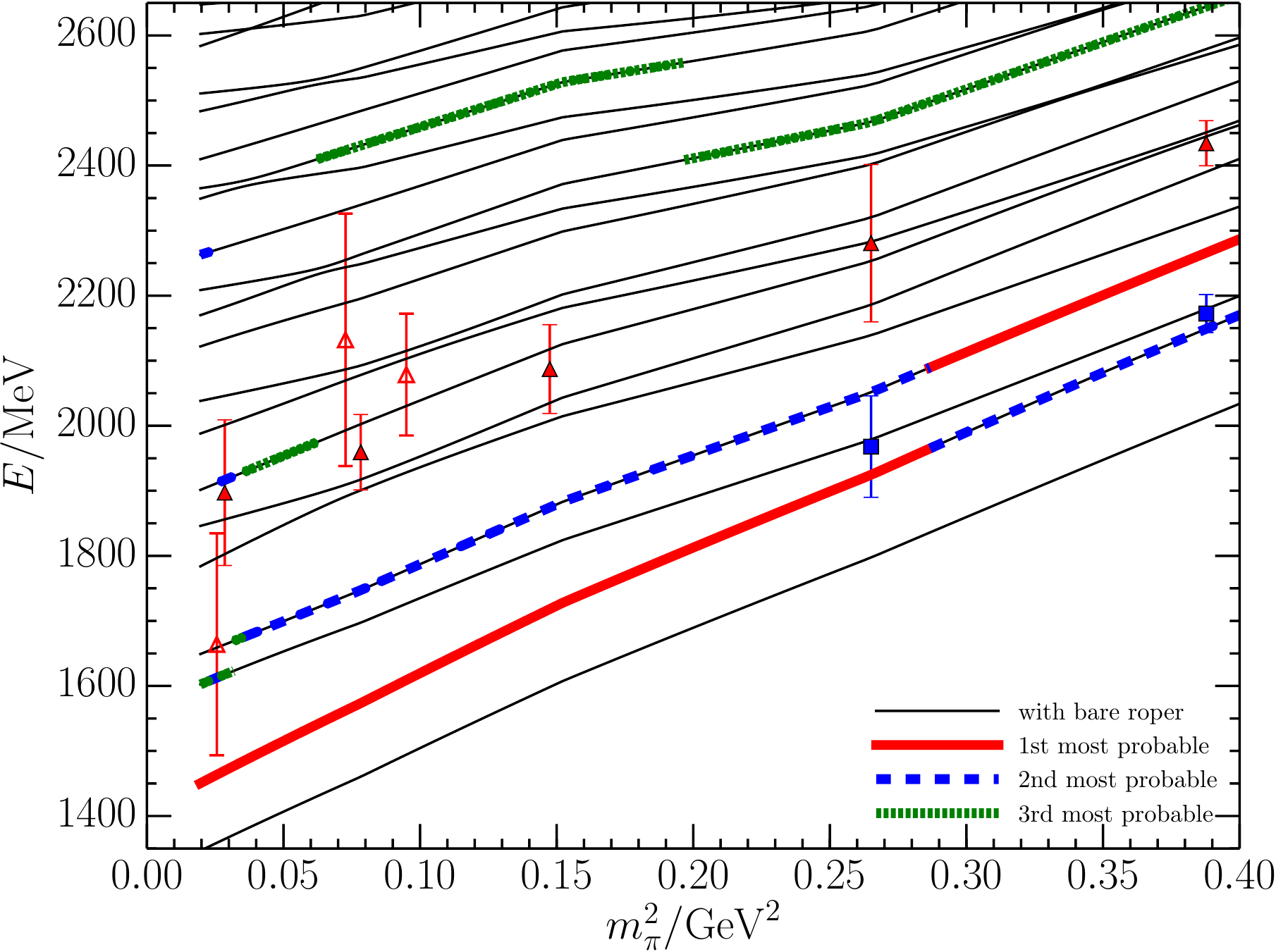}
\includegraphics[width=0.49\columnwidth]{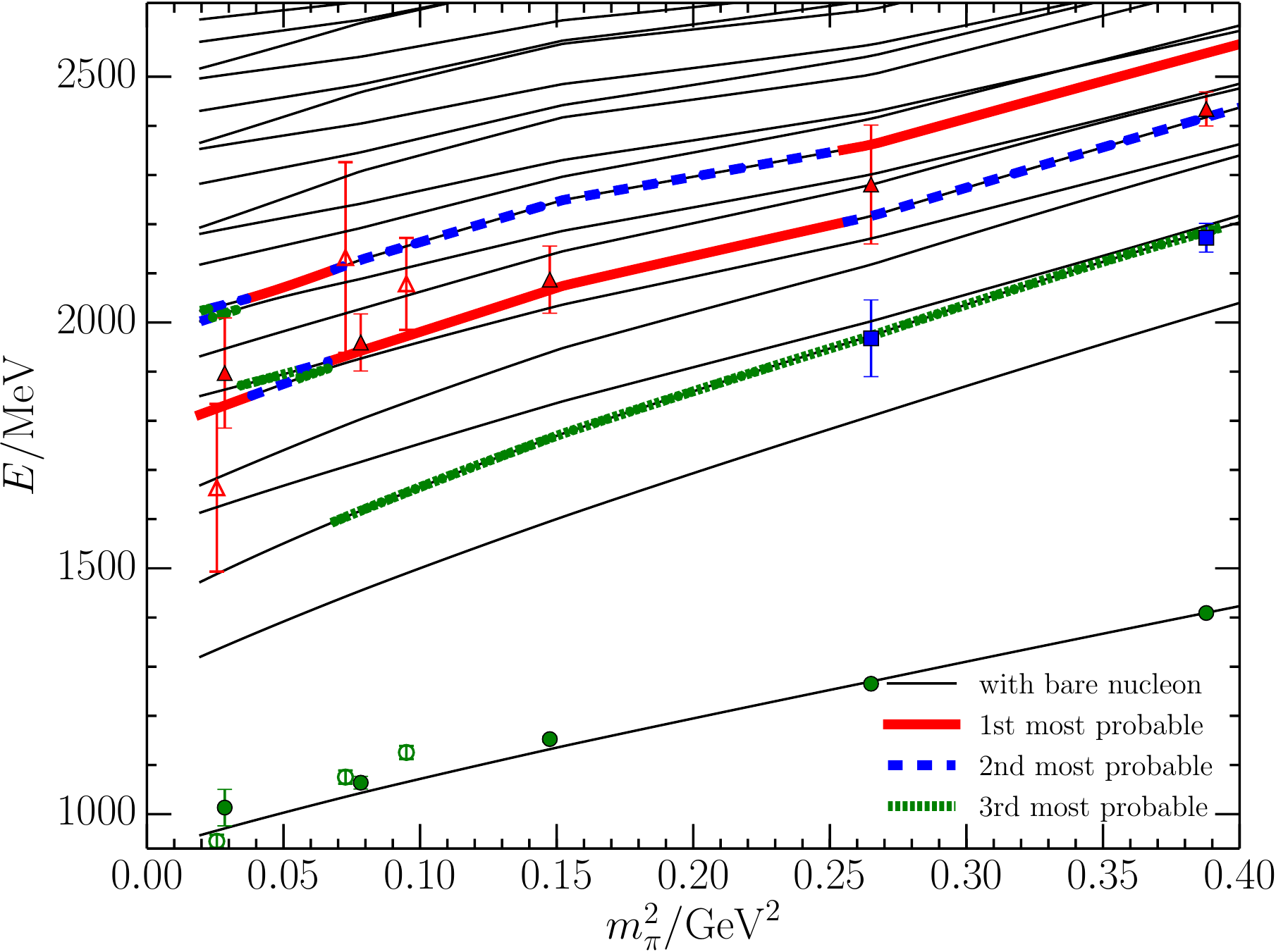}
\caption{{\bf Color online:} The pion mass dependence of the
$L\approx3$ fm finite-volume energy eigenstates for $I(J^P)=\frac12(\frac12^+)$ and $S=0$. The left one is for the scenario with a bare $N^*(1440)$ and the right one is for that with a bare nucleon. See Fig. \ref{fig:N1535Spectrum} for the instructions of line types and colors. }\label{figEnergyLevel}
\end{center} 
\end{figure}

The three scenarios show different behaviors at finite volume. We show the energy levels for the first and third scenarios in Fig.~\ref{figEnergyLevel}. The energy levels for the second scenario are very similar to those for the third, and thus we omit them. From the left graph in Fig. \ref{figEnergyLevel}, we see that the second eigenstate contains about 20\% bare baryon but the lattice simulations do not observe it. This contradiction suggests that the $N^*(1440)$ may contain little or no three-quark component. In the right graph, the lattice QCD data are consistent with the colored lines which represent the most probable states predicted by HEFT. Additionally, we note that there are nontrivial mixings of two-particle states in the eigenstates that overlap with the lattice QCD data.

\section{Numerical results and discussion for $\Lambda(1405)$} \label{SecDisLam1405}

\begin{figure}[t]
\begin{center}
\includegraphics[width=0.49\columnwidth]{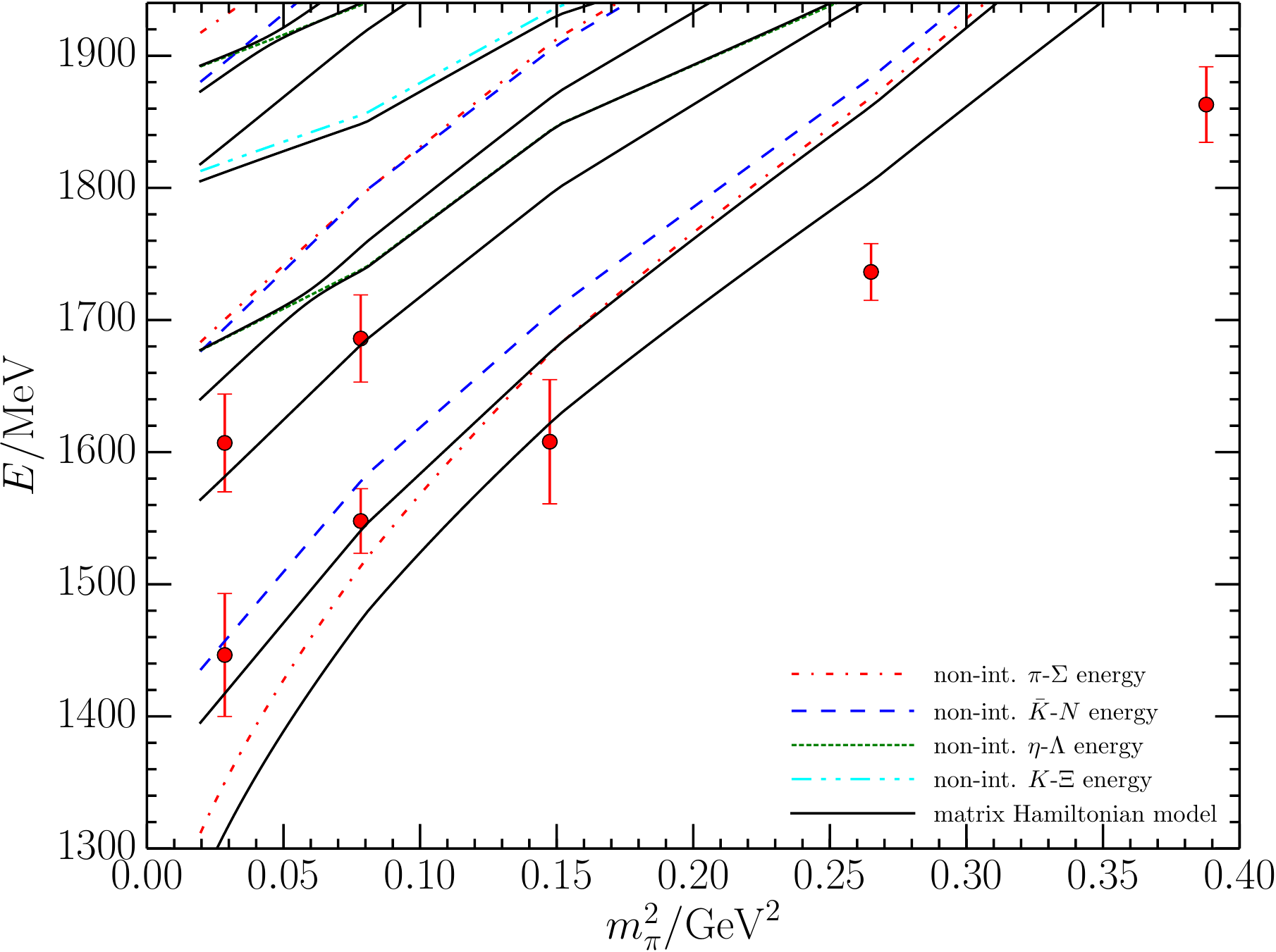}
\includegraphics[width=0.49\columnwidth]{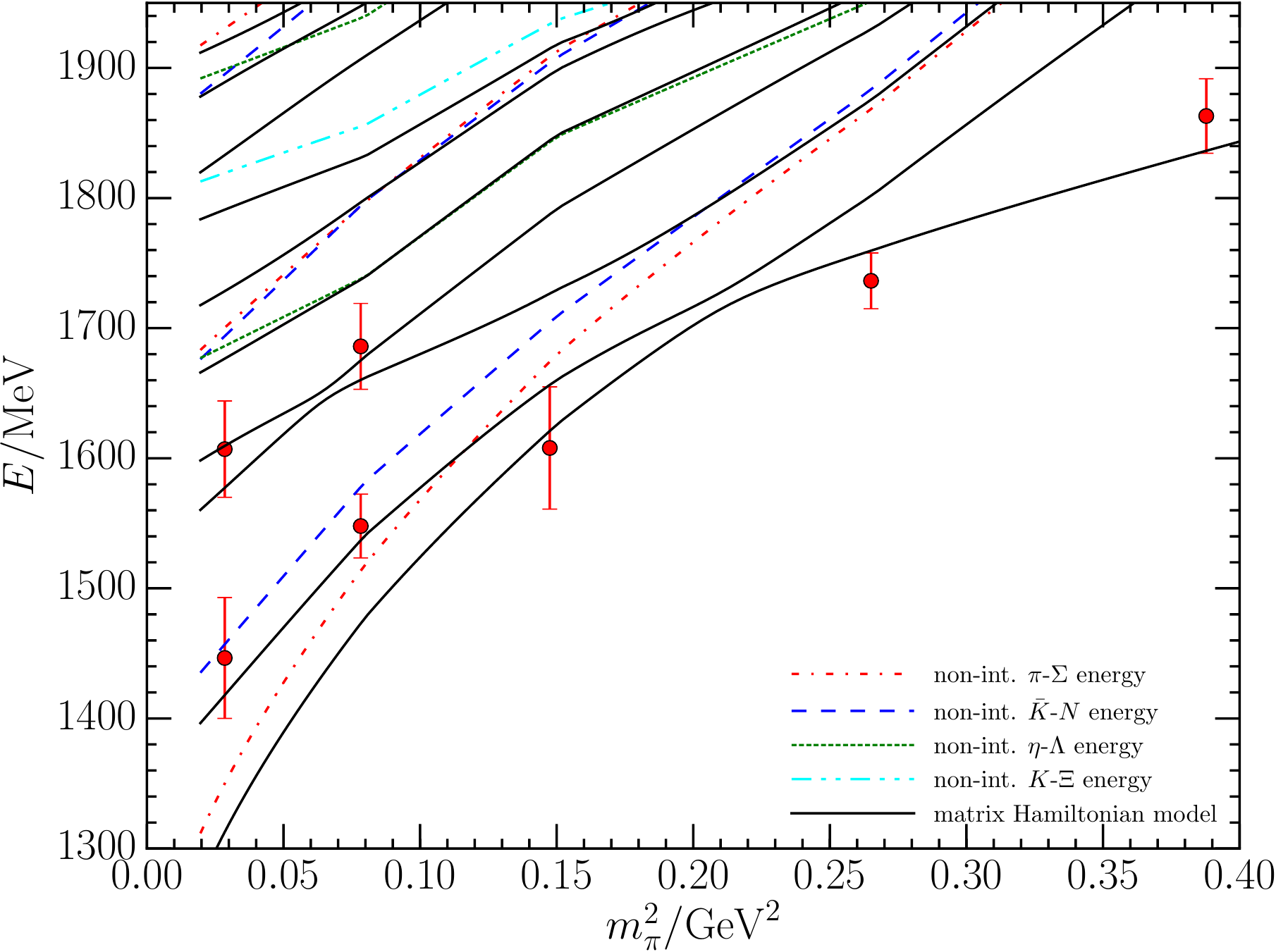}
\caption{{\bf Color online:} The pion-mass dependence of the finite-volume energy eigenstates for the scenarios without (left) and with (right) a bare-baryon basis state for $I(J^P)=\frac12(\frac12^-)$ and $S=-1$. The broken lines represent the non-interacting meson-baryon energies and the solid lines represent the spectrum derived from the matrix Hamiltonian model.  The lattice QCD results are from the CSSM \cite{Hall:2014uca,Menadue2012}.  }\label{figLambda1405}
\end{center} 
\end{figure}

We have studied the cross sections of $K^- p$ and found two poles for $\Lambda(1405)$ at $1430-i22$ MeV and $1338-i89$ MeV. The experimental data can be explained well both with and without a bare baryon. However, the bare baryon component is important for the lattice QCD data at large pion masses. The spectra at finite volume without (left) and with (right) a bare baryon is shown in Fig.~\ref{figLambda1405}. The lattice QCD data at large pion masses is not consistent with the scenario where no bare baryon is considered. There is very little bare baryon in the $\Lambda(1405)$ at small pion masses, but the bare baryon plays an important role at large pion masses.

\section{Summary} \label{SecSummary}
We have studied the $N^*(1535)$, $N^*(1440)$, and $\Lambda(1405)$ with HEFT, analysing both the experimental data and lattice QCD data. The $N^*(1535)$ contains a strong three-quark core while the other two particles do not at the physical pion mass.

\section*{Acknowledgement}
This research is supported by the Australian Research Council through the ARC Centre of Excellence
for Particle Physics at the Terascale (CE110001104), and through Grants No.\ LE160100051,
DP151103101 (A.W.T.), DP150103164, DP120104627 (D.B.L.). One of us (AWT) would also like to
acknowledge discussions with K. Tsushima during visits supported by CNPq, 313800/2014-6, and
400826/2014-3.

\end{document}